\documentclass[11pt,preprint]{aastex}
\usepackage{lscape}
\usepackage{hyperref}
\usepackage{graphicx}
\begin{document}

\title{4DAO Cookbook \\ 
  Version 1.0 \\
  (Nov 2013)
}

\author{Alessio Mucciarelli}
\affil{Dipartimento di Fisica \& Astronomia, Universit\`a 
degli Studi di Bologna,\\ Viale Berti Pichat, 6/2 - 40127
Bologna, ITALY}
\email{alessio.mucciarelli2@unibo.it}

\section{Introduction}

{\tt 4DAO} is a FORTRAN code designed to launch automatically {\tt DAOSPEC} \citep{stetson}
for a large sample of spectra.
The main aims of {\tt 4DAO} are: (1)~to allow an analysis cascade of 
a list of spectra provided in input, by automatically writing the input 
{\tt DAOSPEC} files  and managing its output files; 
(2)~to optimize automatically some spectral parameters used by {\tt DAOSPEC} in the 
process of equivalent width (EW) measurement, above all the Full Width Half Maximum (FWHM); 
(3)~to mask some spectral regions (telluric lines, interstellar features, photospheric 
lines with prominent Lorenztian wings) that can bias the correct EWs measurement;
(4)~to provide suitable graphical tools in order to evaluate the quality
of the solution, especially of the Gaussian fit to each 
individual spectral line; 
(5)~to provide the final normalized, zero radial velocity spectra.


\section{About {\tt DAOSPEC}}
Here the basic use of {\tt DAOSPEC} is drawn but we refer the user to the official documentation 
of this code 
\citep[][and the {\tt DAOSPEC}  Cookbook\footnote{http://www.bo.astro.it/$\sim$pancino/docs/daospec.pdf}]{stetson}. 
{\tt DAOSPEC} is a code that automatically identifies absorption spectral lines, estimates the 
continuum with a Legendre polynomial, measures the EWs and the radial velocities (RVs) 
for all the detected lines, and identifies 
among them the lines provided in an input line list. The measurement of the EWs is performed adopting 
a saturated Gaussian function and using the same FWHM for all the lines. 
One of the most interesting features of {\tt DAOSPEC} is the computation of a global continuum 
that takes into account the effects of weak lines. 

Among the different input parameters, the most important are the value of the FWHM ({\tt FW}), the order 
of Legendre polynomial used to fit the continuum ({\tt ORD}), the residual core flux ({\tt RE}, useful 
to refine the EW measurement for strong lines) and the possible scaling of the FWHM with 
the wavelength ({\tt SC}). 
{\tt DAOSPEC} can run by using a fixed value of the FWHM, chosen by the user, or alternatively the 
FWHM is refined according to the residuals of the spectrum.


\section{Basic layout of {\tt 4DAO}}


At the first run, the analysis starts by adopting as FWHM 
the input value specified by the user and investigating 
a range of RVs specified in input. 
{\tt DAOSPEC} runs, finding new values of FWHM and RV. 
If the difference between the input and output FWHM is larger than 
a threshold value (chosen by the user, see Section 5), a new 
run of {\tt DAOSPEC} is called, starting with the output FWHM of the previous 
run as new input value, and moving in a range between RV-5$\sigma_{RV}$ and 
RV+5$\sigma_{RV}$, where $\sigma_{RV}$ is the dispersion of the mean RV as computed 
by {\tt DAOSPEC} by using the matched lines. During the process of RV determination, 
the {\tt DAOSPEC} parameter {\tt VE} (that is the number of 
standard deviations from the mean radial velocity used to discard the discrepant lines)
is set to 3 by default.

During the first iteration, the {\tt RE} parameter can be tuned 
according to different recipes, chosen by the user (see Section 5). 
Basically, {\tt RE} can be: 

{\sl (i)} chosen by the user as fixed parameter, 

{\sl (ii)} determined by using the central depth of the strongest line 
available along the spectrum. In this case {\tt 4DAO} will read the 
{\tt .daospec} file, looking for the line with the highest EW, among all the 
measured spectral lines (and not only the matched ones). 
Also, in the identification of the strongest line available 
among those measured by {\tt DAOSPEC}, 
{\tt 4DAO} does exclude automatically  
lines lying in spectral regions affected by non-photospheric transitions 
(like telluric and interstellar features). 
These regions are stored in the code and listed in Table 1;

{\sl (iii)} determined by using the central depth of a precise spectral line 
provided by the user.

\begin{figure}[h]
\epsscale{1.0}
\plotone{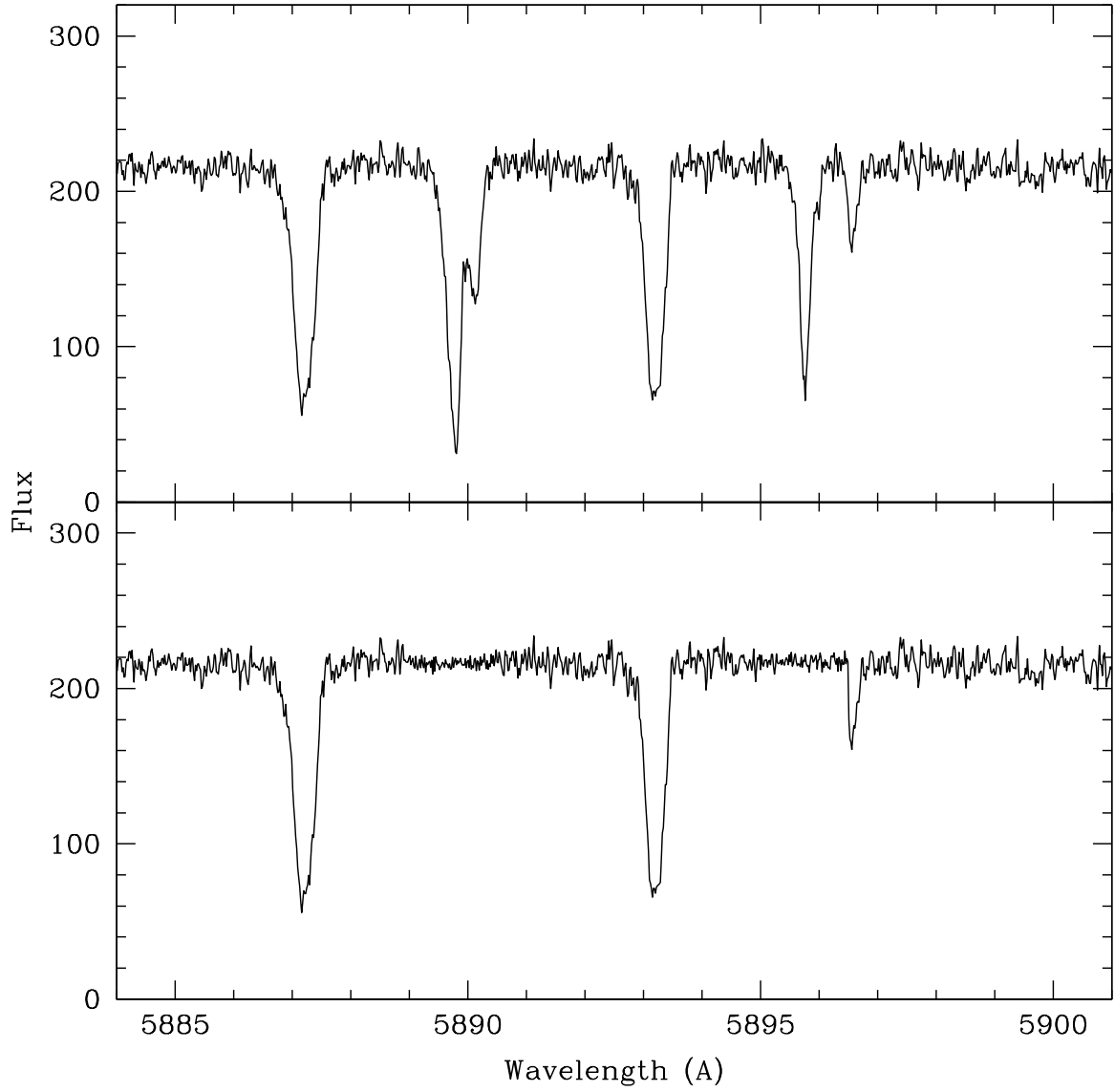}
\caption{Spectral region of the UVES spectrum of the star NGC~5694-37 
around the Na D interstellar lines before (upper panel) and after (lower panel) 
the application of the mask.}
\label{out5}
\end{figure}

The optimization of the FWHM can be affected by the 
presence in the observed spectrum of non-photospheric lines (like telluric 
or interstellar features, whose FWHM will be different with respect 
to that of the photospheric lines), ruined spectral regions or zones 
dominated by wide damped wings (as the Balmer lines or the Calcium II triplet lines). 
{\tt 4DAO} allows to mask spectral regions that the user wish to exclude by the analysis. 
When this option is enabled, {\tt 4DAO} performs a first run of {\tt DAOSPEC}
on the entire spectrum; then the regions to be masked are substituted with the 
continuum spectrum estimated by {\tt DAOSPEC}. 
Because the presence of flat spectral regions can create problems 
with {\tt DAOSPEC}, Gaussian noise (estimated according to the estimated average 
residuals) is added in these regions, 
following the prescriptions by \citet{press}.
Fig.~\ref{out5} shows an example of this process, where the regions corresponding 
to the Na D interstellar lines have been masked in a UVES-FLAMES spectrum 
of a giant star in the globular cluster NGC~5694 \citep{m56}.
This temporary file (called {\tt masked.fits}) 
is used in the following runs of {\tt 4DAO} and when the FWHM converges, 
the last run is performed on the original spectrum.

When the difference between the input and output FWHM reaches the threshold value, 
a final run of {\tt DAOSPEC} is performed, but keeping the FWHM fixed at the 
value optimized in the previous iteration. 
The output on the terminal summarises for each iteration the input FWHM (expressed 
in pixels), the {\tt RE} value, the used range of radial velocities (RV1 and RV2) with 
the average RV  derived in the previous iteration. Additionally, if the spectral mask 
is enabled, a label advises of the use of this option.

\begin{figure}[h]
\epsscale{0.6}
\plotone{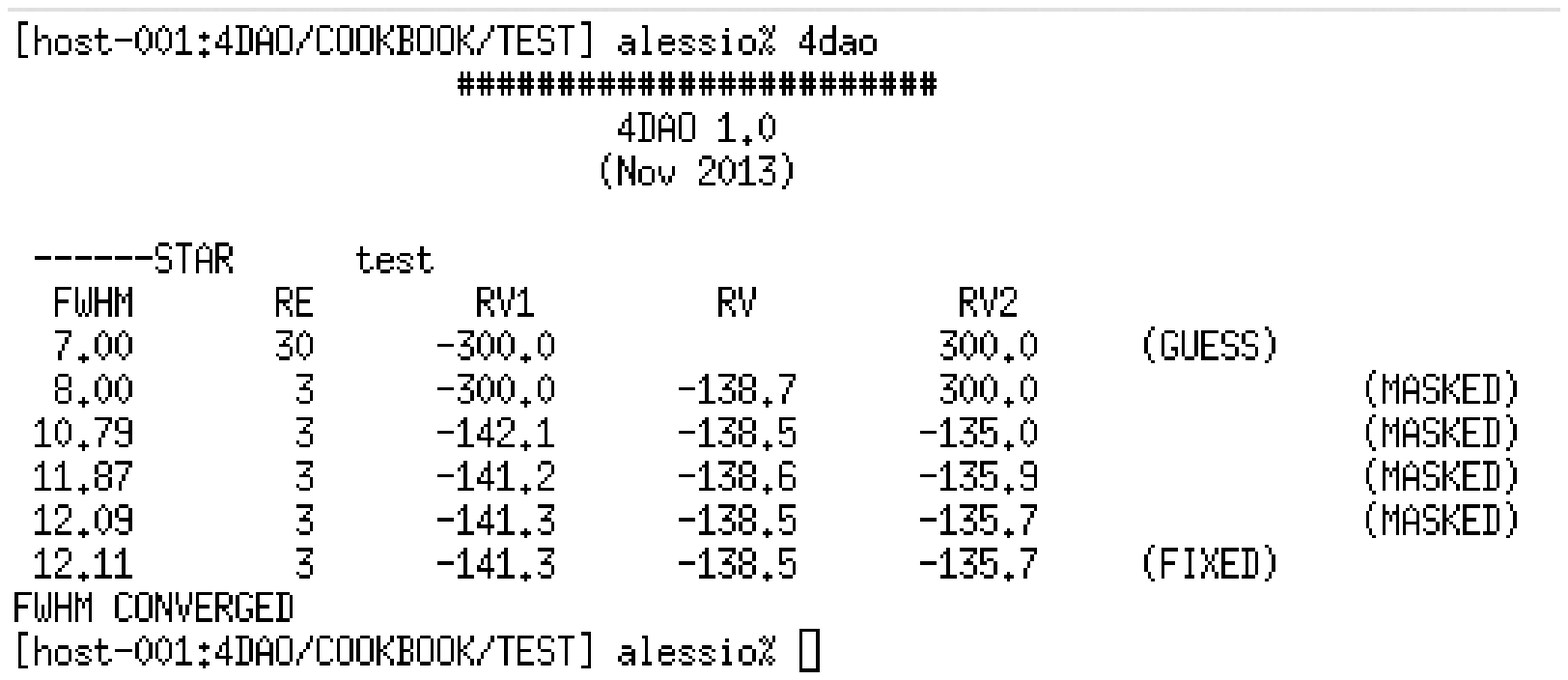}
\end{figure} 

During the execution, the basic input files of {\tt DAOSPEC}, namely 
{\tt laboratory.dat} and {\tt daospec.opt}, are written and managed 
by {\tt 4DAO}, while the output messages that usually {\tt DAOSPEC} 
writes on the terminal, are written in a temporary file named {\tt log.daospec}. 
Also, the graphical output on the monitor usually provided by {\tt DAOSPEC} is 
automatically disabled.

\section{Installation}

Once you have downloaded the archive file {\tt 4DAO\_v*.*.tar.gz} from the website
\begin{center}
\url{www.cosmic-lab.eu/Cosmic-Lab/Products.html},  
\end{center}
type the commands\\
\\
{\tt gunzip 4DAO\_v*.*.tar.gz}\\
{\tt tar -xvf 4dao.tar}\\

These commands unpack the archive file creating a directory named  {\tt 4DAO/}, 
including the source files and the Makefile needed to compile the code. 
Additionally, a sub-directory named {\tt tutorial} includes some examples 
of the configuration files to check quickly if the code is well installed.
To compile the code you need to have installed in your machine the same 
libraries that you have used to install {\tt DAOSPEC} :
(1) the SuperMongo\footnote{http://www.astro.princeton.edu/$\sim$rhl/sm/} (SM) libraries 
(namely {\tt libplotsub.a}, {\tt libdevices.a} and {\tt libutils.a}, compiled in single precision), 
(2) the X11 libraries, and (3) the {\tt libcfitsio.a} library
\footnote{http://heasarc.gsfc.nasa.gov/fitsio/fitsio.html}. 
{\tt 4DAO} can be compiled with the Intel Fortran Compiler, that you can download from 
the Intel website\footnote{http://software.intel.com/en-us/non-commercial-software-development}, 
after the registration.
The {\tt Makefile} can be easily updated, by properly setting the paths of the 
requested libraries. 
Before to start the {\tt 4DAO} installation, check that the installation of all the 
requested libraries is correct (basically, if you have already installed {\tt DAOSPEC} you should be 
already solved all the problems about the installation of these libraries). However, 
we refer the reader to Section 4 of the {\tt DAOSPEC} {\sl Cookbook}
and to Section 3 of the GALA {\sl Cookbook}\footnote{http://www.cosmic-lab.eu/gala/gala.php} 
for the description of some common installation problems and their possible solution. 
Also, {\tt 4DAO} makes use of the {\tt GPL Ghostscript} software (executable {\tt gs}), freely available at 
{\tt http://www.ghostscript.com}; please, check if {\tt gs} is installed on your machine.\\ 
The installation procedure assumes that you have already installed {\tt DAOSPEC} on your machine 
(if you have not yet done, do it!) and that the executable file is named {\tt daospec} (in 
lower-case letters) and its path already stored in your login file. 
If your executable is named with a different name, before to install {\tt 4DAO} you need to 
properly set the variable {\sl nexe} in the {\tt 4dao.f} source file.

Now you can install {\tt 4DAO}, typing the command\\
\\
{\tt make all}\\ 
\\
and the executable {\tt 4dao} will be saved in the current directory. 
Finally, put the path of this directory in your login file according to the shell environment of your 
machine (for instance in the configuration file .bashrc or .tcshrc)

In order to check the installation, go to the {\tt tutorial} subdirectory, hold your 
breath, cross the fingers and type {\tt 4dao}.

\section{Input files}

Only two specific input files (besides the input spectra and the line lists) are necessary 
to run {\tt 4DAO}, a configuration file named {\tt 4dao.param} and a file with the list of 
the spectra to be analysed, named {\tt 4dao.list}. 

(1)~\underline{\tt 4dao.param}\\
The input file {\tt 4dao.param} includes the main configuration parameters, 
adopted in the analysis of all the star listed in {\tt 4dao.list}.
The layout of the file is as follows:

\begin{figure}[h]
\epsscale{0.6}
\plotone{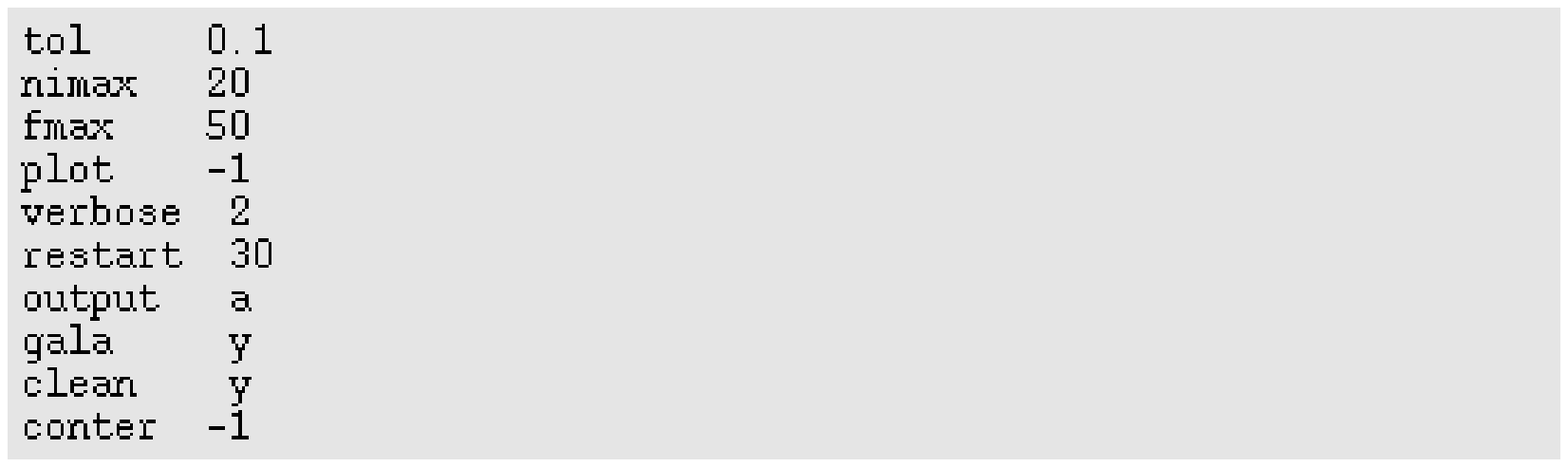}
\end{figure}

\begin{itemize}

\item {\bf tol} is the minimum difference (expressed in pixels) between 
the values of the FWHM derived by {\tt DAOSPEC}  in two consecutive iterations. 
When this difference is smaller than this parameter, {\tt 4DAO} assumes that 
the convergence is reached and the FWHM obtained in the last iteration is 
taken as the final value.
If the FWHM is fixed to the input value (see below for the {\tt 4dao.list} file), 
this parameter is ignored.
\item {\bf nimax} is the maximum number of allowed iterations. Typically, 
{\tt 4DAO} needs of less than 5 iterations to converge to a stable FWHM value, but 
this parameter is useful to avoid infinite loops due to unforeseen problems 
with your spectra. 
\item {\bf fmax} is the maximum value allowed for the FWHM (expressed in
pixels). If the FWHM exceeds the value of this parameter, the convergence process 
is stopped to the last iteration and the star flagged to easily identify the occurrence 
of this problem.
\item {\bf plot} specifies the kind of output plot created for each spectrum. 
The allowed values are 0 (the line plots will be sorted according to their 
wavelengths) and 1 (the line plots will be sorted according to corresponding
element). If different values are provided, {\tt 4DAO} sets automatically this 
parameter to 0.
\item {\bf verbose} specifies the verbosity level on the terminal. 
Accepted values are 0 (no message at all), 1 (only the sequence of the 
analysed spectra is shown), 2 (all the information about the procedure is shown).
\item {\bf restart} is the initial value of the residual core flux {\tt RE} parameter. 
{\tt 4DAO} includes different ways to estimate this parameter, but if the code
fails to derive a reliable parameter (we assumed that {\tt RE} ranges from 0 to 30), 
{\tt RE} will be set to the value specified by {\tt restart}.
\item {\bf output} specifies the format for the output (normalized and 
radial velocity-corrected) spectra: with {\tt F} (or {\tt f}) 
the output spectra will be created in standard FITS format, 
since with {\tt A} (or {\tt a}) will be written in  
ASCII format (the wavelength in the first column and the normalized flux in the second column). 
If different values are provided, the output will be 
created in ASCII format.
\item {\bf clean} deletes some files used by {\tt DAOSPEC}, like 
{\sl laboratory.dat}, {\sl daospec.opt}, {\sl log.daospec} 
and all the files related to the mask of some spectral regions.
The allowed options are {\sl Y} (or {\sl y}) and {\sl N} (or {\sl n}). 
Because in the execution of a sequence of spectra these files are over-written, 
{\tt 4DAO} will save only the files related to the last spectrum. The option {\tt clean}={\sl n}
can be useful to check individually all the temporary files for some problematic spectra.
\item {\bf gala} specifies if the output format is that needed for the input files 
of {\tt GALA} \citep{mgala} or not. 
If the parameter is {\sl Y}/{\sl y} (the default value) the output file will be written 
for {\tt GALA}, for all the other character values, the output file will include all 
the information included in the input line list, reading it as a string.
\item {\bf conter} enables the measurement of the EWs by varying the continuum level. 
This is a crude way to provide a conservative estimate of the impact of the continuum location 
on the measured EWs. During this procedure, the normalized spectrum is lowered and raised by the relative 
flux dispersion in the residual spectrum (as listed in the {\tt .daospec} output file). 
Then, {\tt 4DAO} repeats the same procedure used for the original spectrum, but assuming {\tt ORD}=--1, 
thus fixing the continuum level at 1.
The allowed values are -1 (to disable this option), 0 
(to re-calculate the EWs after a new optimization of the FWHM starting from the best value 
obtained in the main procedure) and 1 
(to re-calculate the EWs by fixing the FWHM at the best value finding in the main procedure). 
\end{itemize}

(2)~\underline{\tt 4dao.list}\\
This file lists the sequence of spectra that you plan to analysis with 
{\tt DAOSPEC}. 
The layout of this file will be as follows:

\begin{figure}[h]
\epsscale{0.6}
\plotone{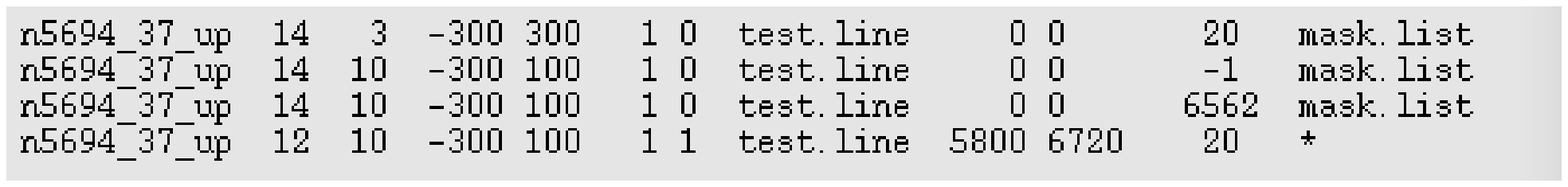}
\end{figure} 

\begin{itemize}
\item The first column indicates the name of the spectrum in FITS format 
(you can also not specify the extension {\sl .fits}).

\item The second and third columns are the initial FWHM (in pixels) and the order of 
the Legendre polynomial, respectively.

\item The forth and fifth columns are the initial range of RV used by 
{\tt DAOSPEC} . 

\item The sixth value is the {\tt DAOSPEC}  parameter {\tt SC}, allowing to enable 
the line fitting procedure assuming that the FWHM is proportional 
to wavelength (see Section 2.3.12 in the {\tt DAOSPEC}  Cookbook). Thus, the 
allowed values are 0 (for the use of a single FWHM for all the lines) and 
1 (for scaling the FWHM according to the wavelength of the lines).

\item The seventh value enables the optimization of the FWHM (0) or 
launch {\tt DAOSPEC}  keeping the FWHM fixed to the value specifies in the second column 
of the file (1).

\item The eighth column specifies the line list used for that spectrum. 
No specific format is requested, but only that the first column is 
the wavelength in $\mathring{A}$ and the second the code of the element. 
For the latter, {\tt 4DAO} accepts both the {\tt GALA} format (i.e. 26.00 for 
Fe~I and 26.01 for Fe~II) and the {\tt MOOG} format (i.e. 26.0 for Fe~I and 
for 26.1 for Fe~II). Note that if the keyword {\tt gala} in {\tt 4dao.param} 
is {\sl Y}, the input file needs to include all the information requested by {\sl GALA} 
\citep[wavelength, element code, log~gf, excitation potential, damping constants and 
$\alpha$ velocity parameter, see][]{mgala}:

\begin{figure}[h]
\epsscale{0.6}
\plotone{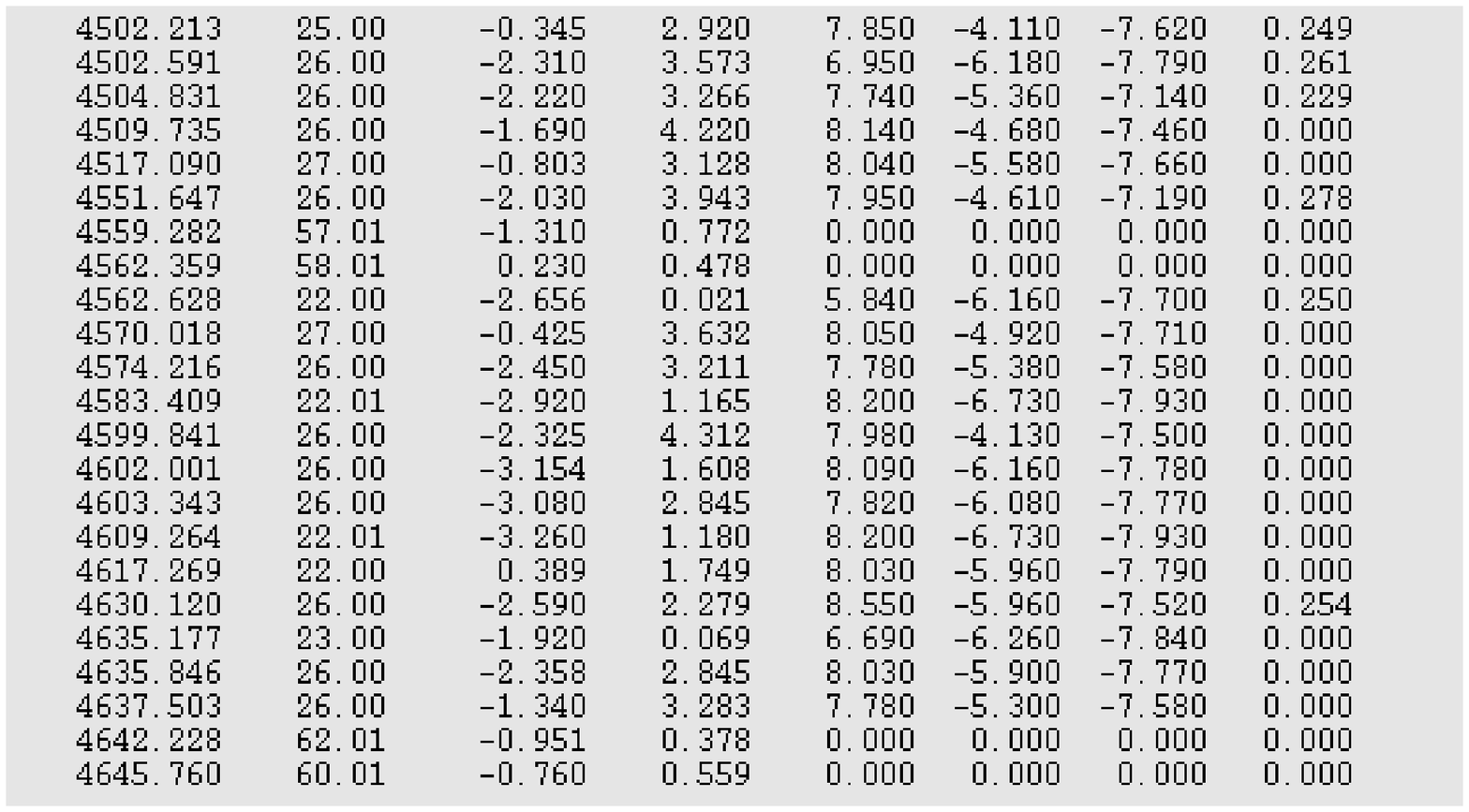}
\end{figure}

\item The ninth and tenth values are the wavelengths that identify 
the spectral range where {\tt DAOSPEC}  performs the spectral line measurements. 
You can specify  the spectral range that you prefer; if one (or both) of the 
value is set to 0, {\tt 4DAO} tries to readjust the corresponding spectral 
edge in order to avoid regions with negative flux, too noisy or with 
dramatic variations of the flux with respect to the spectrum.

\item The eleventh column specifies the way to set the {\tt RE} parameter. 
If a value between 0 and 30 is provided, {\tt RE} will be fixed at this value.
For negative values, {\tt RE} is fixed to the value specified in {\tt 4dao.param} 
by the keyword {\tt restart} 
for the first iteration, then it is refined by using the central depth 
of the strongest line measured by {\tt DAOSPEC} .
Alternatively, you can provide the wavelength of a given strong line 
(for instance a Balmer line, if available) and {\tt 4DAO} will try to match its position 
with the closest transition measured by {\tt DAOSPEC} .
In both cases, if the derived {\tt RE} values is negative or larger 
than 30, the value specified by {\tt restart} will be assumed.

\item The last column is the name of the ASCII file including the wavelengths 
of spectral regions that you want to mask during the analysis. 
The file does not need any specific format, but it 
includes only two columns (each raw corresponding to a given spectral region 
to be masked, considering the observed spectrum, thus the wavelengths do not 
include the RV shift.).
If you do not need to use this option, you can only specify a name or a symbol 
(like * in the some raws of the shown example) not corresponding to an existing file.
\end{itemize}

\section{Output files}

\begin{figure}[h]
\epsscale{0.9}
\plotone{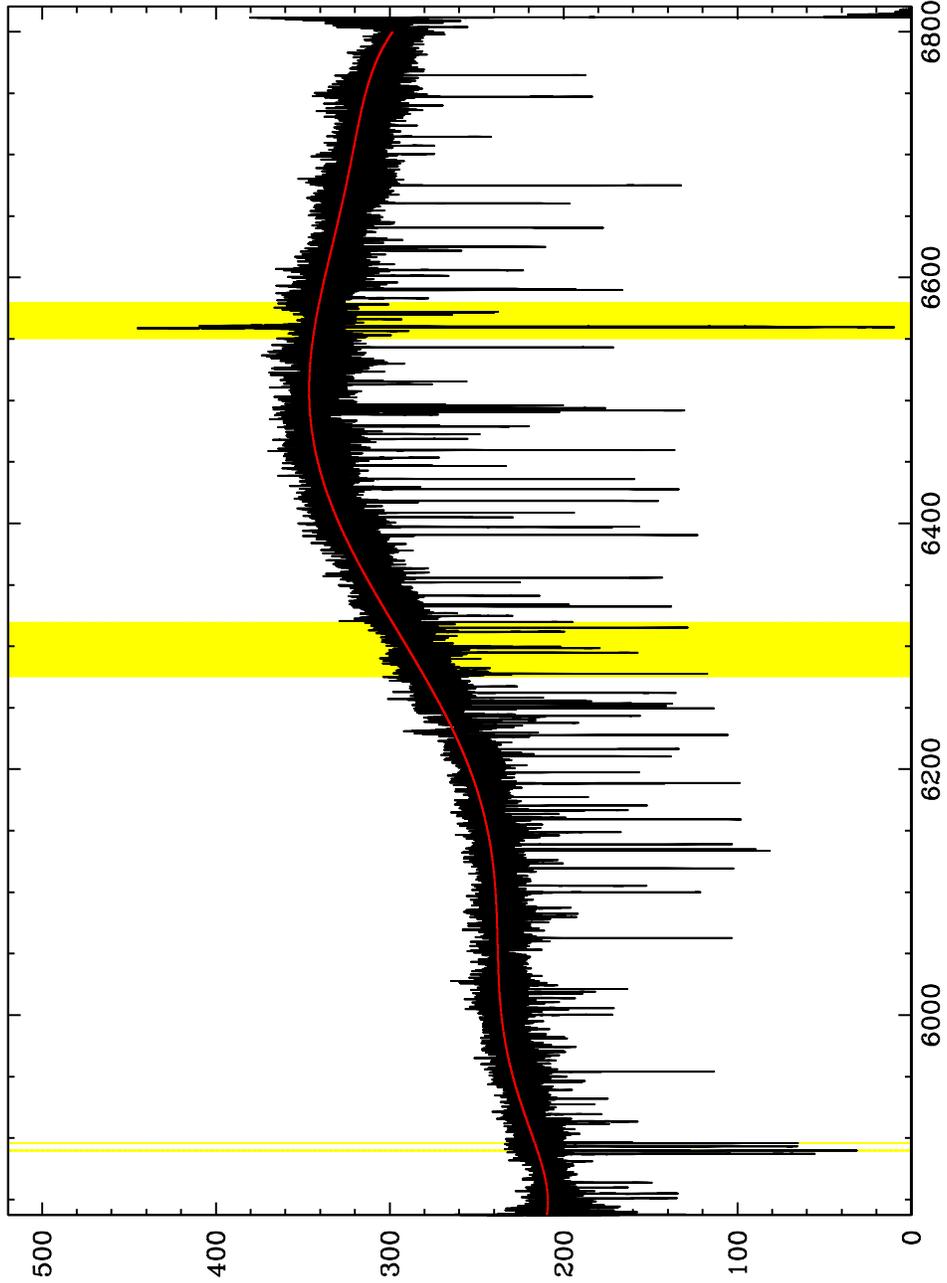}
\caption{The UVES Red Arm 580 spectrum of the star NGC~5694-37 \citep{m56} 
with superimposed the continuum computed by {\tt DAOSPEC} (red curve). The 
yellow-shaded areas mark the spectral regions masked by {\tt 4DAO}, namely 
the Na D photospheric lines, the region contaminated by telluric lines between 
6280 and 6320 $\AA$ and the region around the $H_{\alpha}$ Balmer line.
}
\label{out1}
\end{figure}

\begin{figure}[h]
\epsscale{1.0}
\plotone{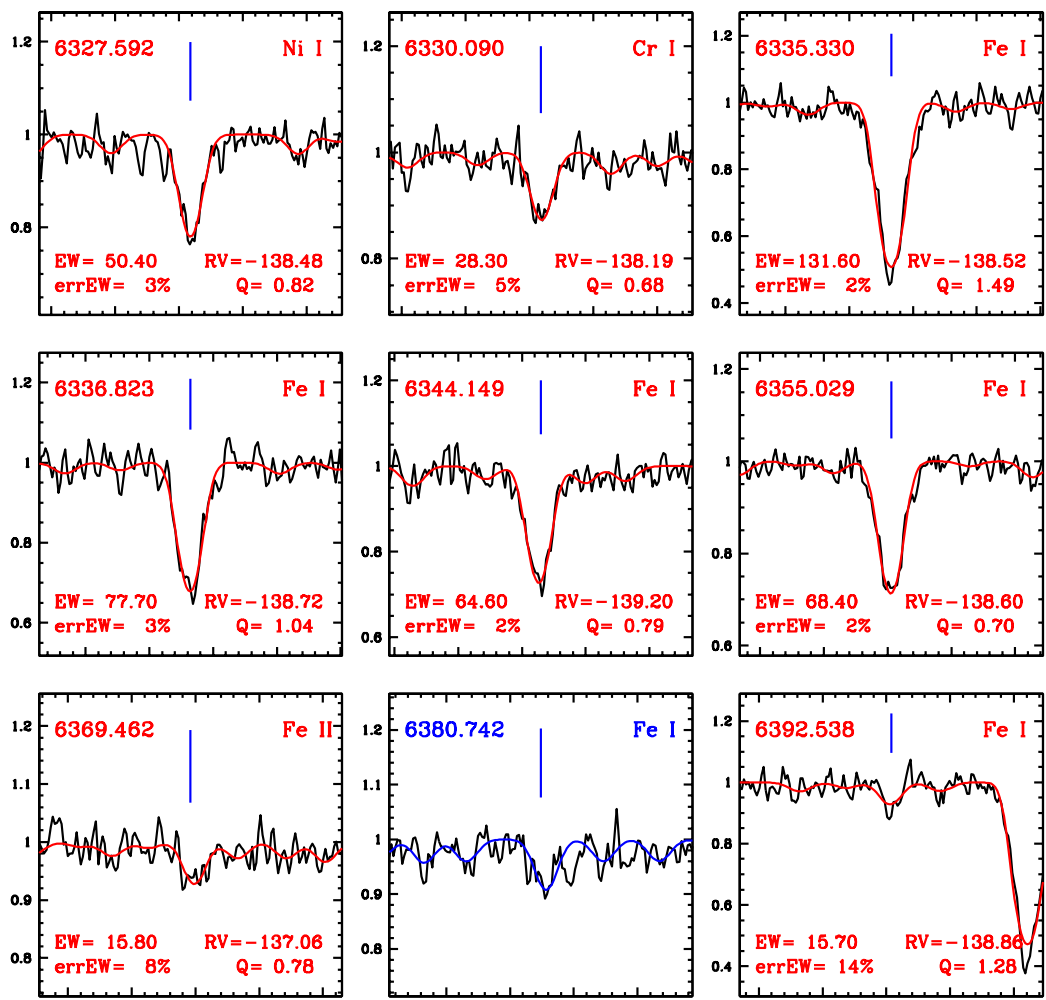}
\caption{Example of the {\tt 4DAO} plots showing the fit (red lines) 
of each individual line; blue lines mark the lines not matched by {\tt DAOSPEC}. 
In each sub-panel the main information about the line (wavelength, ion code, 
EW and its error, radial velocity and Q parameter) are listed.
}
\label{out2}
\end{figure}

\begin{figure}[h]
\epsscale{1.0}
\plotone{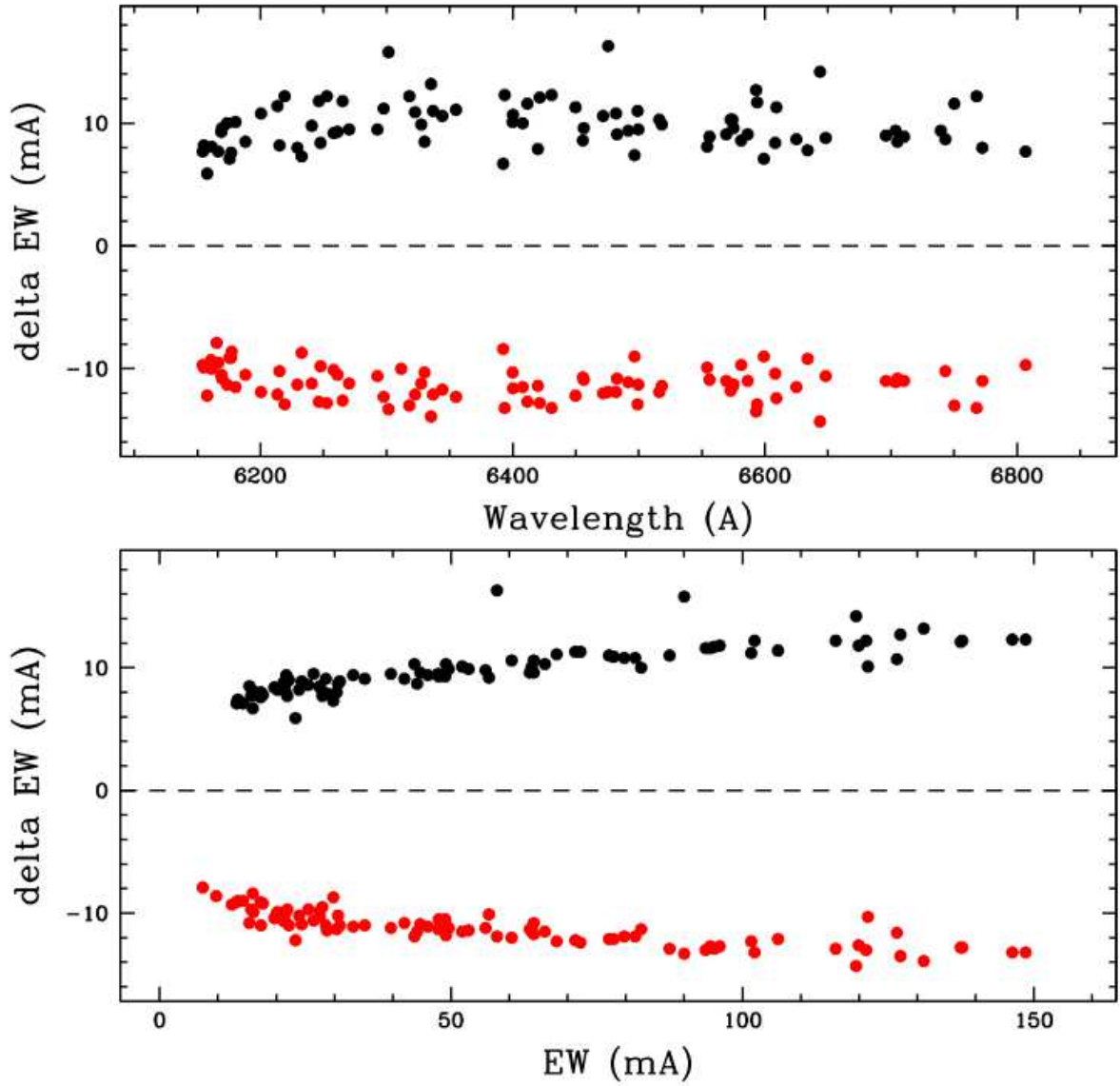}
\caption{Variations of the measured EWs calculated increasing (black points) 
or decreasing (red points) the continuum level according to the flux residuals.
These variations are shown as a function of the wavelength (upper panel) 
and of the EW (lower panel).
}
\label{outc}
\end{figure}

\begin{figure}[h]
\epsscale{1.0}
\plotone{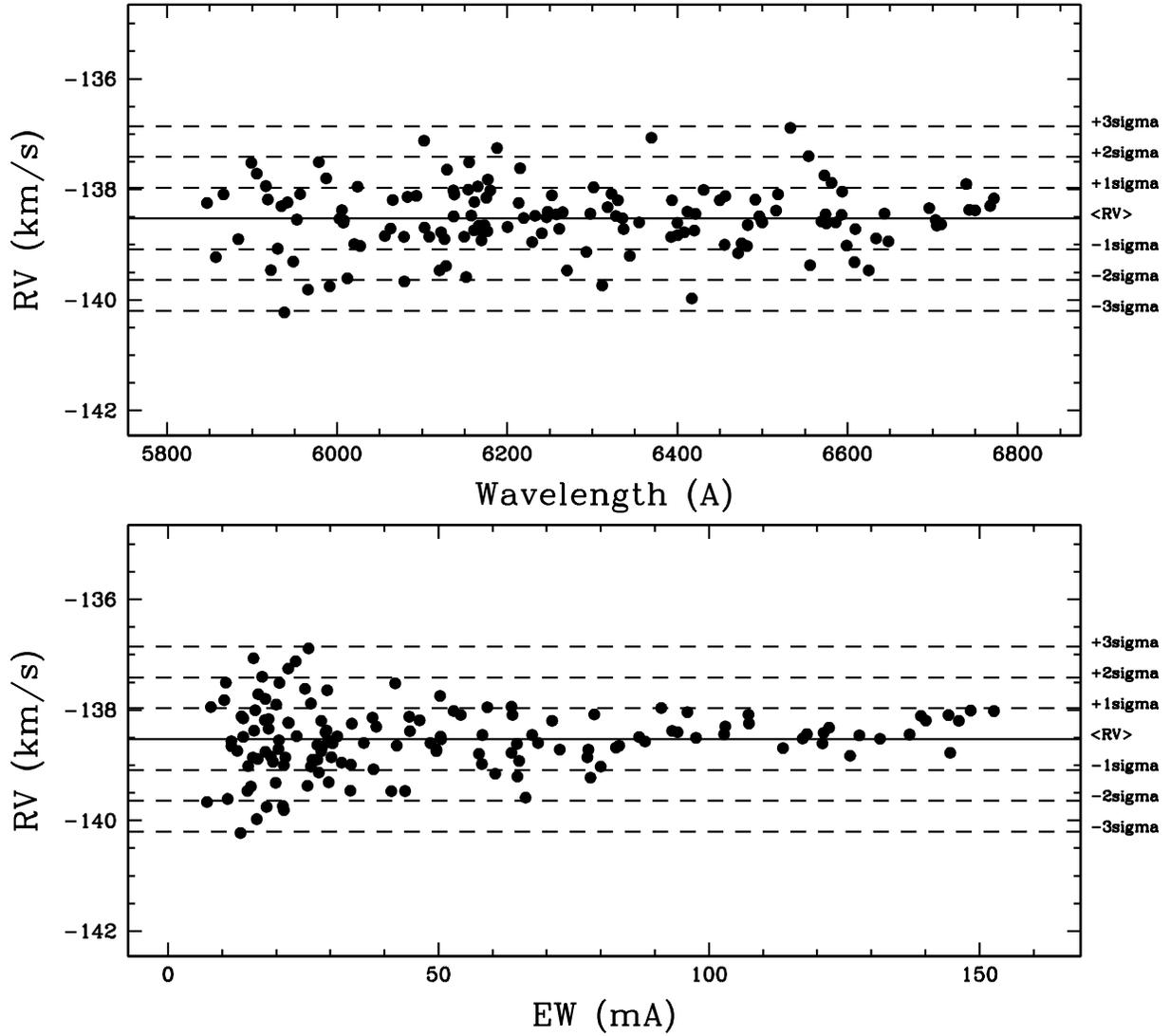}
\caption{Behavior of the radial velocity of each individual 
spectral line as a function of the EW (upper panel) and of the 
wavelength (lower panel). In both the panels the dashed horizontal 
lines are $\pm$1$\sigma$, $\pm$2$\sigma$ and $\pm$3$\sigma$ levels.
}
\label{out3}
\end{figure}

\begin{figure}[h]
\epsscale{1.0}
\plotone{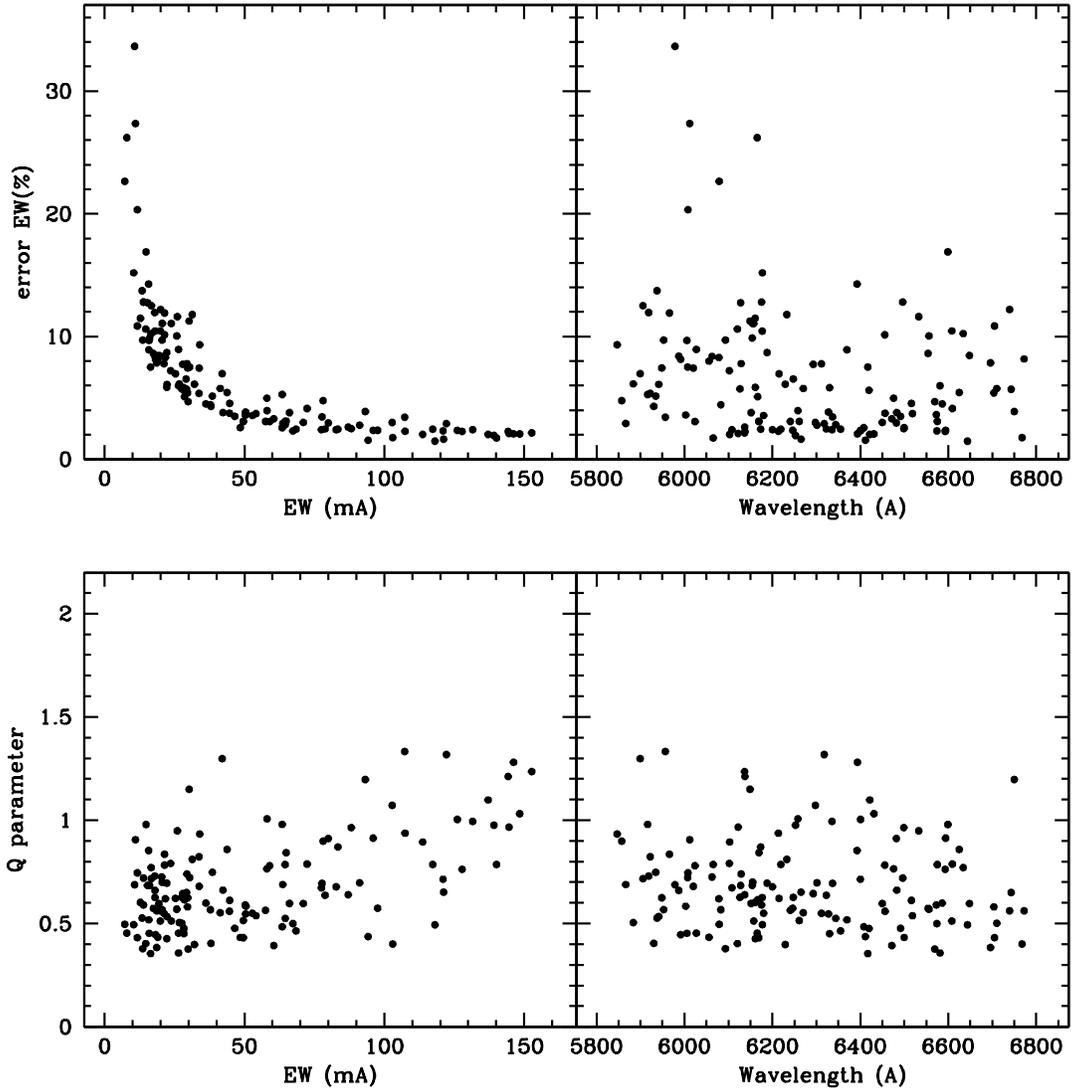}
\caption{Upper panels: behavior of the EW error (expressed in percentage) 
as a function of the EW and of the wavelength. Lower panels: 
behavior of the Q parameter as a function of the EW and of the wavelength.
}
\label{out4}
\end{figure}

Together with the standard output files produced by {\tt DAOSPEC}  
(the FITS files including the fitted continuum and the 
residual spectrum, and the {\tt .daospec} file with all the 
measured lines), {\tt 4DAO} produces some output files to check 
the quality of the solution and manage the derived information.
For the spectrum named {\tt rootname} (as specified in the first column 
of {\tt 4dao.list}), the following files 
are created:
\begin{itemize}
\item {\tt rootname}\_4DAO.pdf includes some plots concerning the 
continuum derived by {\tt DAOSPEC} , the fit of each individual line and 
information about RVs and EWs uncertainties.

The first panel shows the entire spectrum with superimposed (as a red line)
the continuum level computed by {\tt DAOSPEC} . If you have masked some spectral 
zones, these regions will be shown in this plot as yellow-shaded regions 
(see Fig.~\ref{out1}). The RV shift of the star is not applied 
in this plot.

The following panels display all the lines listed in the 
input file, sorted in the wavelength or in the element code 
(according to the keyword {\tt plot} in {\tt4dao.param}), with superimposed the 
best-fit (red line) calculated by {\tt DAOSPEC} . Lines in the input file 
that are rejected or not recognized by {\tt DAOSPEC}  are plotted in 
blue color, in order to allow an easy identification of the 
lost features. In each panel the main information are labelled, 
as wavelength, ion code, EW, radial velocity, uncertainty in EW 
(expressed in percentage) and Q parameter. An example of these plots is shown 
in Fig.~\ref{out2}.

The panel shown in Fig.~\ref{outc} is created only if the keyword {\tt conter} is 
0 or 1. It shows the variation of the measured EWs with respect to the original values 
when an increase or a decrease of the continuum level is assumed (black and red points, 
respectively). The variation of EWs is shown as a function of the wavelength (upper panel) 
and of the EW (lower panel).

The second to last panel (see Fig.~\ref{out3}) shows the RV of all the lines as a function 
of the wavelength (upper panel) and the EW (lower panel), and with the 
$\pm$1$\sigma$, $\pm$2$\sigma$ and $\pm$3$\sigma$ levels marked as dotted lines.

The last panel (Fig.~\ref{out4}) shows the behavior of the EW error (upper panels) 
and of the Q parameter (lower panels) as a function of EW and wavelength.

\item the file named {\tt rootname.in} includes the main information about the EW measurements. 
If the {\tt GALA} output is enabled, this file will have the same format described 
in the {\tt GALA} Cookbook, with the addition of the Q-parameter (not requested by {\tt GALA}) 
in the eleventh column.
Alternatively, it will contain wavelength, EW, error in EW, Q-parameter and then 
all the other information provided in the input line list.
Note that if {\tt conter} is 0 or 1, two additional columns will be added (at the end of the file), 
including the EWs measured by raising and lowering the normalized spectrum, respectively.

\item {\tt rootname}\_ZVN (.fits or .dat according 
to your choice in the keyword {\tt output} in {\tt 4dao.param}) 
is the original input spectrum, normalized using the continuum 
calculated by {\tt DAOSPEC} 
and corrected for radial velocity using the average RV derived 
by {\tt DAOSPEC} . This file is especially useful to create scientific plots 
or to use to perform additional chemical analysis based on the spectral 
synthesis.
\end{itemize}

Additionally, the file {\tt daospec.log} summarises the main parameters derived 
by {\tt 4DAO} for all the spectra listed in {\tt 4dao.list}. 
For each spectrum the final FWHM (in pixels), the average radial 
velocities (in km/s) with its dispersion, the number of matched lines, the 
flux residuals (in percentage), a convergence flag related to the FWHM, the used starting and ending wavelengths, 
the {\tt RE} value and the wavelength of the line used to derived {\tt RE} (only in case this parameter 
is tuned by using the strongest available line) are provided. 
The file header explains 
the meaning of the convergence flag. Briefly:\\ 
CONV=1 the FWHM does converge (or the FWHM has been fixed to the initial value without optimization);\\
CONV=0 the number of iterations exceeds the maximum number of allowed iterations (specified by {\tt nimax} 
in {\tt 4dao.param}). In this case the results are referred to the last iteration;\\
CONV=--1 if the code calculates the same value of FWHM in two different iterations,
to avoid the risk of an infinite loop, the code exits, writing the results of the last iteration, 
and passing to the next spectrum;\\ 
CONV=--2  the FWHM exceeds its maximum allowed value specified by {\tt fmax} in {\tt 4dao.param}
Also in this case, the written values in the output are referred to the last iteration. 
This flag identifies also the cases where a negative FWHM is provided or found;\\
CONV=--3 the median value of the entire spectrum is negative, pointing out some problems in the 
spectral reduction (or the spectrum is missing). 
{\tt DAOSPEC}  would crash with similar spectra, thus the analysis is stopped, all the 
output values are 0.0 and {\tt 4DAO} moves to the next spectrum of the list\\
CONV=--4 format problems in the {\tt daospec.opt} or {\tt .daospec} files are found;\\
CONV=--5 no line is found in the wavelength range of the observed spectrum.


\begin{deluxetable}{lll}
\tablecolumns{3} 
\tablewidth{0pc}  
\tablehead{\colhead{Feature} & \colhead{$\lambda_{start}$}& \colhead{$\lambda_{end}$} \\ 
\colhead{} & \colhead{($\mathring{A}$)} & \colhead{($\mathring{A}$)}}
\startdata 
\hline	   
Na~D first component  &   5889.0  &	5890.5      \\ 
Na~D second component &   5895.0  &	5896.5      \\ 
$O_2$ X0-b2 Band    &	6275.0  &  6320.0   \\ 
B Band	    &	6860.0  &  6930.0    \\ 
$H_2$O Band   &	7160.0  &  7330.0      \\ 
A Band	    &	7590.0  &  7700.0      \\ 
$H_2$O Band   &	8125.0  &  8340.0      \\ 
$H_2$O Band   &	9100.0  &  9800.0      \\ 
 &     & 	  \\ 
\hline
\enddata 
\tablecomments{Spectral regions masked by {\tt 4DAO} during the determination 
of the {\tt RE} parameter.}
\end{deluxetable}

\end{document}